\documentclass[12pt,a4paper]{article}
\usepackage[T1]{fontenc}
\usepackage{lmodern,amsmath,amssymb,amsthm,bm,mathtools}
\usepackage[a4paper,textwidth=170mm,textheight=672pt,centering]{geometry}
\usepackage{setspace,graphicx,booktabs,array,natbib,microtype}
\usepackage[font=normalsize,labelfont=bf,skip=5pt]{caption}
\usepackage[hidelinks]{hyperref}
\usepackage{newfloat,float}
\DeclareFloatingEnvironment[fileext=loa,name=Algorithm]{algorithm}
\floatstyle{ruled}
\restylefloat{algorithm}
\usepackage{enumitem}
\usepackage{subcaption}

\usepackage{titlesec}
\titlespacing*{\section}{0pt}{14pt plus 2pt minus 4pt}{6pt plus 1pt minus 3pt}
\titlespacing*{\subsection}{0pt}{10pt plus 2pt minus 4pt}{4pt plus 1pt minus 3pt}
\newtheorem{theorem}{Theorem}
\newtheorem{proposition}{Proposition}

\newcommand{\E}{\mathbb E}
\newcommand{\Var}{\operatorname{Var}}
\newcommand{\KL}{\operatorname{KL}}
\newcommand{\diag}{\operatorname{diag}}
\newcommand{\tr}{\operatorname{tr}}
\newcommand{\R}{\mathbb R}
\newcommand{\ind}{\mathbf 1}
\newcommand{\pos}[1]{(#1)_+}
\newcommand{\argmax}{\operatorname*{arg\,max}}
\newcommand{\argmin}{\operatorname*{arg\,min}}
\newcommand{\RMSE}{\operatorname{RMSE}}
\newcommand{\Rel}{\operatorname{Rel}}

\usepackage{color}

\title{Nonnegativity Shapes Predictive Uncertainty:\\ Complementary Model Fits and Quantile Construction}
\author{Jialin Li\\[-2pt]
\small University of Massachusetts Amherst, Amherst, MA, USA
\quad \texttt{jialinl@umass.edu}}
\date{}
\makeatletter
\renewcommand{\maketitle}{\begin{center}{\large\bfseries\@title\par}\ifx\@author\@empty\else\vspace{0.1em}{\normalsize\@author\par}\fi\end{center}\vspace{0.1em}}
\makeatother
\renewenvironment{abstract}{\begin{center}\textbf{Abstract}\end{center}\begin{quote}\normalsize}{\end{quote}}
\begin{document}
\maketitle
\begin{abstract}
Knowing that a response cannot be negative can inform predictive uncertainty, including extreme upper quantiles. This knowledge matters especially when a black-box system returns long vector responses at few covariate settings, leaving large uncertainty across many entries. We develop a framework that assembles predictive distributions feature by feature from distributional fits that incorporate nonnegativity, optionally around any given mean predictor. The fits come from a latent factor model that represents response intensity through nonnegative patterns whose amplitudes follow rectified Gaussian processes over covariates. A variational algorithm jointly learns the factorization and the latent posterior. Varying a conditional-dispersion parameter yields a spectrum of fits that show complementary strengths in estimating distributional shape features. An orthogonal quantile expansion motivates using L-moments or their ratios as features. We select a fit separately for each using scores computable from a single validation observation per entry, then assemble the selected estimates through the expansion. In simulations, this construction reduces extreme upper-quantile errors by 17–22\% relative to the respective best single fit, despite the mode-seeking tendency of our variational inference. In a streamflow demonstration across agricultural catchments, using an external point simulation to supply the mean reduces miss rates of rare shortage by 86--91\%.

\end{abstract}
\noindent\textbf{Keywords:} Extreme quantiles; Gaussian process; L-moments; nonnegative matrix factorization; predictive distribution; variational inference.

\section{Introduction}\label{sec:intro}
Accurate predictive distributions of weather-dependent outcomes help farms select crops resilient to volatile rainfall and streamflow, lenders and insurers price solar-farming contracts, and hospitals plan capacity before heat waves. However, these outcomes can carry large uncertainty, for two reasons. First, the underlying weather conditions are changing. Record-shattering climate extremes are projected to become more probable \citep{fischer2021increasing}, and daily precipitation variability has increased over more than 75\% of land with observational coverage during the past century \citep{zhang2024anthropogenic}. Second, the systems of interest are black boxes. They map covariates such as weather conditions to a response whose distribution changes with the covariates in ways that are poorly understood. For example, vegetation responds to rainfall frequency and intensity even after accounting for annual totals \citep{feldman2024large}. Together, these two uncertainty sources pose modeling challenges. Errors in predictive distributions, especially in predicting rare shortage or exceedance that can drive a decision, can propagate into substantial downstream consequences.

An accurate predictive distribution must respect what is known about the response. Outcomes such as crop yields, rainfall, streamflow, and hospital admissions cannot be negative. Gaussian predictions is nonetheless common for nonnegative data (e.g., \citealt{yu2008gaussian}), and large uncertainty can lead to substantial violations of this bound. A common remedy clips predictions at zero after fitting, but the distribution above zero remains unchanged. A parametric model may guarantee nonnegativity but rely on assumptions that are hard to justify when extreme events become frequent. Where a stochastic simulator exists, simulating a response distribution at one covariate setting requires many expensive repeated runs. Predicting under untested conditions or predicting extreme events can add extra error and cost. These difficulties compound when the response of interest is a vector. The rainfall evidence above shows that annual summary variability can mislead and daily rainfall uncertainty supports better planning for a new agricultural season. More generally, outputs indexed by time, location, or label can reveal distribution heterogeneity. However, more outputs also give a bound-ignoring model more opportunities to violate it.

In such scenarios, uncertainty quantification can acknowledge this type of prior knowledge for higher accuracy, and let the bound inform predictive distributions. This paper develops a framework that takes into account nonnegativity in model design, handles vector responses, predicts response distributions at any covariate settings, and aims for both accurate extreme quantiles and low integrated errors. Rather than requiring one model to describe every shape feature of a quantile function, the framework assembles it feature by feature from complementary model fits that each respect the bound. The framework is semiparametric and data-driven, providing flexibility to fit to a black box while requiring no known mechanism of the system. It can accommodate predictive uncertainty around any existing mean predictor or simulator.

Specifically, we consider a black-box system that returns $Y(\bm x)\in[0,\infty)^T$ at a covariate setting $\bm x\in\R^D$. Each output index $t$ refers to a fixed response entry and need not have a meaningful ordering. We suppose that nearby covariate settings generate similar response distributions, but assume no smoothness across output indices (e.g., rainfall uncertainty can change substantially across days). Observing a new setting may be costly or time-consuming, so the number of observed settings $N$ can be modest while $T$ is large. Our hydrology demonstration (Section~\ref{sec:demo}) illustrates this structure with daily runoff across 42 catchments, given four weather covariates. Each seasonal response has $T=3{,}780$ catchment-day entries, with only $N=172$ observed seasons. Other functional data may share this structure as well.

Motivated by nonnegative matrix factorization (NMF), we first propose a nonnegative latent Gaussian process (GP) factor model, with response initially considered as Poisson (Section~\ref{sec:model}). The model represents response intensities through shared normalized nonnegative output patterns, while the rectified GP describes how their amplitudes change with the covariates and accommodate inactive patterns over part of the domain. Learned latent coordinates in the GP kernel encode dependence among GP components. The factorization and the GP are fitted jointly, so the predictability of the amplitudes can inform a generally nonunique factorization.

We derive a variational algorithm for joint fitting (Section~\ref{sec:inference}), named EMEM, from the integrated probabilistic view of the model, inspired by that of generalized Kullback--Leibler (KL) NMF  (Section~\ref{sec:prelim}). We smooth the amplitude distribution near zero during fitting, making the marginal likelihood (i.e., evidence) finite without explicitly inferring the zero pattern. The evidence lower bound then guides EMEM's four alternating updates for component allocations, factorization, the latent Gaussian posterior, and GP hyperparameters. The allocations and factorization updates avoid high-dimensional numerical optimization while the latent Gaussian approximation has dimension independent of $T$. A mean-preserving jittered Poisson model extends the same updates to continuous responses.

We extend the observation model with a parameter $G$ that controls conditional dispersion while preserving the conditional mean structure (Section~\ref{sec:dispersion}). It generates a spectrum of models with the same allocation and factorization update formulas, but changes the balance between response reconstruction and compatibility with the GP prior. We show that a large-$G$ limit connects near-optimal joint fits and separate fitting. As $G\downarrow 0$ integrated quantile error diverges when the fitted mean remains bounded away from zero. Numerical evidence shows that different fits between these $G$-spectrum ends can capture different distributional features most accurately, as measured by L-moment ratios \citep{hosking1990}. This motivates using the model family as a source of complementary shape estimates for quantile construction.

We construct predictive quantile functions from L-moments, or their ratios, each selected separately across candidate fits and combined through an established expansion (Section~\ref{sec:quantile}). A reconstruction error bound expresses quantile error in terms of errors in separate moments and truncation, justifying separate moment selection. The first L-moment is the mean, so an external predictor/simulator can provide it while the candidate fits provide dispersion and higher-order shape. To select moments when each $(\bm x,t)$ entry has only one observation, we specialize the influence-function-adjusted responses of \citet{chetverikov2025} to L-moments and use them to rank candidate moments or ratios. We show the corresponding reference-bias and response-sensitivity bounds. The bounds motivate selecting a common reference by the continuous ranked probability score \citep{gneiting2007} within nested validation.

In synthetic experiments where responses follow a distribution outside every candidate family (Section~\ref{sec:synthetic_exp_design}), the construction with ratio selection reduces quantile and distribution-function integrated squared errors by 31.1\% and 19.4\% relative to the respective best single-$G$ candidate. At probabilities 0.99, 0.999, and 0.9999, its quantile RMSE is 16.7\% to 22.1\% lower than that of the respective best candidate. Diagnostics identify mean prediction as a major source of remaining integrated quantile error and a truncation-accuracy tradeoff in construction for extreme-tail errors. In the hydrology demonstration (Section~\ref{sec:demo}), supplying the mean from an existing streamflow simulation reduces lower-miss rates by 86.0--91.4\% and interval scores by 12.2--22.5\%, and closes on average 81.5\% of the gap between observed and nominal coverage.

These results suggest several methodological insights. Uncertainty near the lower bound motivates the model design. When lower-tail predictions remain close to zero, upper-tail errors can strongly influence L-moment estimation errors and guide selection. Knowledge of the lower bound thus shapes upper-tail uncertainty quantification in this framework. Although reverse-KL variational inference is usually considered mode-seeking, the fitted predictive distributions can still provide useful complementary information for response-tail prediction through quantile construction. An accurate external predictor or simulator can reduce the ``first-order'' contribution to quantile error. More broadly, the mean and the uncertainty at every order can be estimated by separate models specifically built for that purpose.

\paragraph{Related work.}
A central idea of our framework is to assemble a predictive distribution from features selected separately across fitted candidates. Related methods characterize conditional distributions through individual quantiles \citep{koenker1978}, or through location, scale, and shape parameters modeled by separate covariate functions \citep{rigby2005}. Other approaches combine forecasts into predictive distributions. Statistical postprocessing uses a Gaussian with ensemble-dependent parameters \citep{gneiting2005} or a mixture centered on bias-corrected forecasts \citep{raftery2005}. Bayesian stacking learns mixture weights through cross-validated predictive scores \citep{yao2018}. Quantile aggregation allows weights to vary across probability levels \citep{berrisch2023}, while more flexible constructions also allow covariate-dependent weights and combinations across input quantile levels \citep{fakoor2023}. Our construction selects L-moments or their ratios separately by order and combines the resulting moments through an orthogonal quantile expansion \citep{hosking1990}.

Latent GP factor models in this paper provide a flexible source of candidate distributions for this construction. \citet{higdon2008} combine dimension reduction with GP emulation of basis coefficients, showing similar model structure. Other models share information across outputs by linearly mixing latent GPs \citep{teh2005}, assigning GP priors to both loadings and factors \citep{luttinen2009}, or convolving shared latent processes \citep{alvarez2011}. GP regression networks further allow input-dependent mixing weights \citep{wilson2012}. Physical structure can also inform probabilistic prediction through conditional generative models trained using governing equations \citep{zhu2019}. Constrained GP methods can encode positivity and other inequalities \citep{swiler2020}. More directly related to our factorization, \citet{schmidt2008} link nonnegative matrix factors to GP priors through strictly increasing transformations. Our model incorporates nonnegativity through jointly learned nonnegative output patterns and rectified GP amplitudes, allowing components to be inactive without requiring governing equations.

\section{Base model and motivation}\label{sec:model}

Let $Y=(y_{tn})\in[0,\infty)^{T\times N}$ denote the response vectors, with column $n$ observed at a covariate setting $\bm x_n\in\R^D$. Fitting $T$ separate models or one model over $t$ may scale badly in $T$.
A factor representation instead expresses each response vector through a smaller collection of common, standardized output patterns, whose amplitudes vary with $\bm x$. We therefore consider learning such pattern representation from data, with Gaussian process (GP) used to capture amplitude similarity across covariate settings and account for predictive uncertainty.

An established approach constructs a basis by singular value decomposition (SVD) and models its components using GPs \citep{higdon2008}. This results in Gaussian predictive distributions generally assign positive probability to negative responses. Clipping predictions at zero enforces nonnegativity after fitting but implies a change in the predictive probability model, which is not incorporated in estimation and inference.

\begin{figure}[!htbp]
\centering
\begin{subfigure}[t]{.48\linewidth}
\centering
\includegraphics[width=\linewidth]{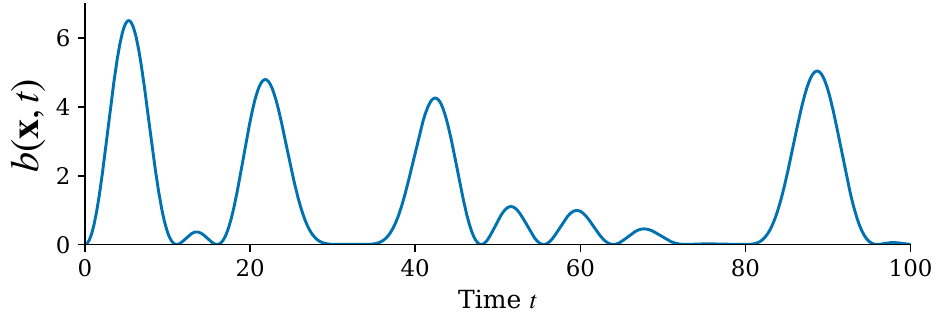}
\caption{Base function.}\label{fig:base}
\end{subfigure}\hfill
\begin{subfigure}[t]{.48\linewidth}
\centering
\includegraphics[width=\linewidth]{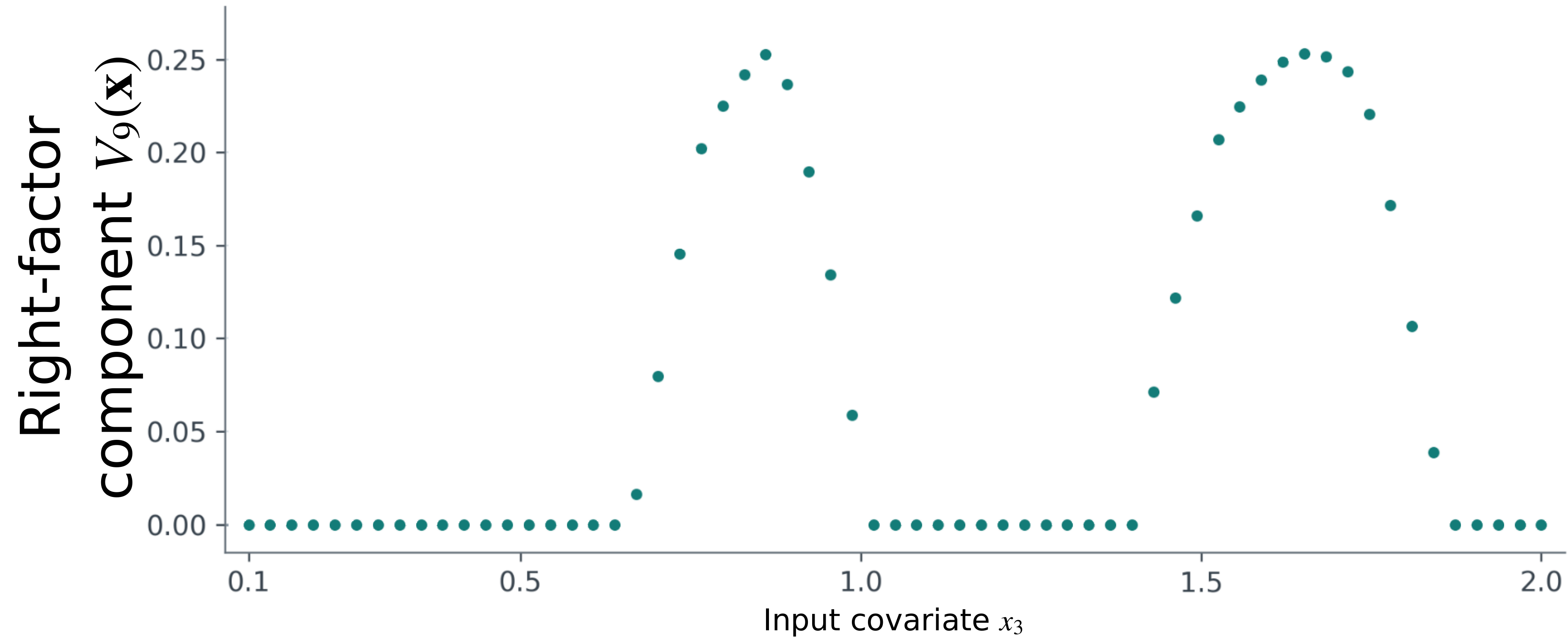}
\caption{One right-factor component.}\label{fig:nmf}
\end{subfigure}
\caption{(a) The base function \eqref{eq:base} used for numerical experiment setup, along $t$ with $\bm x=(1.512,1.964,1.885)$. Its peaks and near-zero stretches move across covariate settings. (b) One row of the right factor $V$ from a rank-ten NMF of the base function \eqref{eq:base} over a grid, along $x_3$ with $x_1=x_2=1.05$. Broad exact zeros motivate a rectified Gaussian process.}
\label{fig:base-nmf}
\end{figure}

A natural nonnegative analogue first computes a nonnegative matrix factorization (NMF), $Y\approx W\Lambda V$ with $W$ and $\Lambda$ normalized, and then fits GPs to the rows of the right factor $V$. NMF represents observations as additive combinations of nonnegative patterns, as illustrated by its early use in learning parts of images \citep{lee1999}. Figure~\ref{fig:nmf} shows a related feature in our problem setup that a right-factor component is active over some parts of the covariate domain and exactly zero over others. An ordinary nondegenerate GP neither respects nonnegativity nor assigns positive probability to an exact zero. A rectified GP, obtained by taking the positive part, accommodates both positive values and inactive regions.

Rectification alone does not resolve all modeling challenges. NMF is also generally non-unique. A separate GP fit must accept the factors given by the first NMF step. We therefore treat the GPs as latent and formulate the factorization and GPs jointly, so that the GPs inform the most likely factors. This motivates the following nonnegative factor model and our integrated inference problem.

We model a component amplitude at a covariate setting $\bm x$ and its response intensity at $t$ by
\begin{equation}\label{eq:model}
V_p(\bm x)=\pos{f(\bm x,\bm z_p)+\varepsilon_p(\bm x)},\qquad \varepsilon_p(\bm x)\sim N(0,\sigma_v^2),\qquad \nu_t(\bm x)=\sum_{p=1}^P w_{tp}\lambda_pV_p(\bm x),
\end{equation}
where $a_+=\max(a,0)$, $w_{tp}\ge0$, $\sum_tw_{tp}=1$, $\lambda_p\ge0$ and $\sum_p\lambda_p=1$. In analogy to NMF $Y\approx W\Lambda V$, the left factor $W=(w_{tp})$ contains $P$ output pattern vectors (representing the pattern in, e.g., Figure \ref{fig:base}), $\Lambda = \diag\{(\lambda_p)\}$ contains component weights, and $\bm V(\bm x) = (V_p(\bm x))$ component amplitudes shared across the output coordinates. Both $W$ and $\Lambda$ are treated as factorization parameters to learn. The amplitude noise $\varepsilon_p$ is independent across components and settings.
Initially, take $Y_t(\bm x)\mid\bm V(\bm x)\sim\operatorname{Poisson}\{\nu_t(\bm x)\}$ independently across $t$, so the intensity $\nu_t$ is the conditional mean. This Poisson model is motivated by the probabilistic view of NMF, which we review in Section \ref{sec:prelim}. Section~\ref{sec:continuous} extends it to a continuous model with the same conditional mean. As a map from latent Gaussian values to $\nu$, the conditional mean \eqref{eq:model} presents a neural network structure with one rectified hidden layer and nonnegative output weights.

In particular, NMF does not impose the orthogonality available in a SVD, for which independent GPs are reasonable. To learn dependence between components, we follow \citet{zhang2020latent} and assign a latent coordinate $\bm z_p\in\R^{d_z}$ to each component and use one zero-mean GP on the augmented domain $(\bm x,\bm z)$ with self-determined dimension $d_z$. We use an elementary kernel

\begin{equation}\label{eq:kernel}
k\{(\bm x,\bm z),(\bm x',\bm z')\}=\sigma_f^2\exp\Big\{-\frac12\sum_{d=1}^{D}\frac{(x_d-x_d')^2}{\ell_d^2}-\frac12\|\bm z-\bm z'\|^2\Big\},
\end{equation}
with length scales $\ell_d$ and common variance $\sigma_f^2$. The length scale for $\bm z$ is absorbed into the inferred coordinates $\bm z$, and nearby coordinates give strongly correlated components. This kernel design defines one GP combining all components with homoscedasticity in $\bm x$ and heteroscedasticity in $\bm z$, while the framework is open to more complicated kernels. Stacking $f_{pn}:=f(\bm x_n,\bm z_p)$ component by component into vector $\bm f\in\R^{PN}$ gives the prior covariance
\begin{equation}\label{eq:kronecker}
K_\Theta=\sigma_f^2(K_z\otimes K_x),\qquad \Theta=(\bm z_1,\ldots,\bm z_P,\ell_1,\ldots,\ell_D,\sigma_f^2)
\end{equation}
in the Kronecker product form, where $K_x$ and $K_z$ are respective correlation matrices over settings and coordinates. Compared with the original \citet{zhang2020latent}, the adopted Kronecker form ensures no increase in the order of magnitude of covariance computation cost with domain augmentation. $\Theta$ denotes learnable hyperparameters. Because $K_\Theta$ is invariant under shifting and rotations of all $\bm z_p$ together, for $d_z=2$ and $P\ge2$ we fix $\bm z_1= \bm0$ and the first entry of $\bm z_2$ at zero for identifiability, leaving $2P-3$ coordinate entries to learn (we reuse notation $\Theta$ with fixed entries removed).

Amplitudes $V = (V_p(\bm x_n)) \in [0,\infty)^{P\times N}$ are conditionally independent given latent GP $\bm f$, and output responses are conditionally independent given $V$. Marginally, shared latent GP makes responses dependent, so the model structure allows us to learn even if data is made of only one sample from each unique $(t, \bm x)$ combination.

We seek to estimate the model parameters $(W, \Lambda, \Theta)$ by maximizing the marginal likelihood of $Y$, also called the evidence, and to infer the posterior distribution of the latent function $p(f|Y)$. Joint fitting allows the latent GP to favor more compatible factorizations among those giving similar reconstructions. Evaluating the evidence, however, requires integration over the latent variables. We therefore use variational inference, jointly optimizing the model parameters and an approximate posterior through an evidence lower bound (ELBO) on the log marginal likelihood \citep{blei2017}. Our choice of the generalized Kullback--Leibler NMF (KL-NMF) objective, rather than another reconstruction loss, is intentional for deriving the ELBO. KL-NMF admits a Poisson likelihood interpretation that simplifies the variational inference and yields explicit updates for the factorization parameters (see Section~\ref{sec:inference}). Section~\ref{sec:prelim} reviews this view of KL-NMF and its frequent zeros. A further difficulty comes from the unknown zero pattern of $V$. Section~\ref{sec:inference} begins by explaining the issue and addresses it by smoothing.

\section{The probabilistic view of KL-NMF}\label{sec:prelim}
We briefly follow \citet[Section 2]{cemgil2009} for the two-factor NMF $Y \approx WV$. For nonnegative $W = (w_{tp})\in[0,\infty)^{T\times P}$ and $V= (v_{pn})\in[0,\infty)^{P\times N}$ with $\nu_{tn}=(WV)_{tn}$, independent Poisson observations $Y$ have log likelihood $\sum_{tn}\{y_{tn}\log\nu_{tn}-\nu_{tn}-\log(y_{tn}!)\}$, so maximum likelihood minimizes the generalized KL loss $D(Y\Vert WV)=\sum_{tn}\{y_{tn}\log(y_{tn}/\nu_{tn})-y_{tn}+\nu_{tn}\}$, which is defined for all nonnegative data. We do not consider least-squares NMF because its objective implies independent Gaussian errors \citep{schmidt2008} which do not impose nonnegative support. The key step writes each observation as a sum of latent Poisson component counts,
\begin{equation}\label{eq:superposition}
H_{tpn}\mid W,V\ \overset{\rm indep.}{\sim}\ \operatorname{Poisson}(w_{tp}v_{pn}),\qquad Y_{tn}=\sum_{p=1}^P H_{tpn}.
\end{equation}
The superposition property of Poisson random variable enables us to write down the likelihood of $Y$. The posterior of component counts at an output entry is multinomial with ``total'' $y_{tn}$:
\begin{equation}\label{eq:classical-allocation}
H_{t\cdot n}\mid Y,W,V\sim\operatorname{Multinomial}(y_{tn},r_{t\cdot n}),\qquad r_{tpn} :=\frac{w_{tp}v_{pn}}{\nu_{tn}},\quad h_{tpn}:=\E H_{tpn}=y_{tn}r_{tpn}.
\end{equation}
Each observation is allocated in proportion to the intensity each component contributes. For fixed expected allocations, the part of the complete log likelihood depending on the factors is $\sum_{tpn}\{h_{tpn}\log(w_{tp}v_{pn})-w_{tp}v_{pn}\}$. Holding the other factor fixed, its maximization gives updates $w_{tp}\leftarrow {\sum_nh_{tpn}}/{\sum_nv_{pn}}$ and $v_{pn}\leftarrow {\sum_th_{tpn}}/{\sum_tw_{tp}}$,
which with \eqref{eq:classical-allocation} are the multiplicative updates of \citet{lee2001}. Alternating these steps with refreshed allocations gives the EM interpretation of KL-NMF \citep{cemgil2009}.

With a diagonal $\Lambda=\diag(\lambda_1,\ldots,\lambda_P)$ in NMF, $w_{tp}v_{pn}$ is replaced by $w_{tp}\lambda_pv_{pn}$. Normalizing the columns of $W$, the rows of $V$ and $\Lambda$ gives the multinomial mixture model of \citet{kubjas2015}, and absorbing the total into $V$ gives our design \eqref{eq:model} (see Supplement~S1.1). Zero entries in a factor are common as seen in the random-matrix experiments of \citet[Table 1 and Example 3.4]{kubjas2015}. Likelihood maxima often lie on the boundary of the nonnegative-rank model even for strictly positive data. Such observation supports our rectified GP model.

\section{Variational inference}\label{sec:inference}

\subsection{The unknown zero pattern}\label{sec:smoothing}
Writing $\phi$ and $\Phi$ for the standard Gaussian density and distribution function, each amplitude $V_p(\bm x_n)$ has a conditional probability mass of size $\Phi(-f_{pn}/\sigma_v)$ at zero and a density on $(0,\infty)$. If the zero pattern of $V$ were known, the likelihood would multiply probability masses for zero entries by densities for positive ones. With the pattern unknown, the likelihood scale is inconsistent. As a second issue, any positive allocation to a component gives expected log likelihood $-\infty$ because $\E\log V_p(\bm x_n)=-\infty$  (Section \ref{sec:prelim} presents similar terms).

We therefore replace the zero mass and a small neighborhood by a conditional uniform variable during fitting. For small fixed $\delta>0$, suppose $\zeta_{pn}\mid f_{pn}\sim N(f_{pn},\sigma_v^2)$, model $V_p(\bm x_n)=\zeta_{pn}$ if $\zeta_{pn}>\delta$, and otherwise model $V_p(\bm x_n)$ as a uniform variable on $(0,\delta)$. In general notation,
\begin{equation}\label{eq:smoothing}
p_\delta(v\mid f)=\frac{\Phi((\delta-f)/\sigma_v)}{\delta}\ind_{(0,\delta)}(v)+\frac1{\sigma_v}\phi((v-f)/\sigma_v)\ind_{(\delta,\infty)}(v).
\end{equation}
The uniform part has finite expected log value $\log\delta-1$, so allocation now needs no explicit inference or conditioning of the zero pattern. This smoothed model is used only for fitting with prediction still using \eqref{eq:model}.
The smoothing does not make a large impact for other $V_p(\bm x_n)$ related terms in the ELBO as we show later. In fact, using the same $\zeta_{pn}$, $|V_p(\bm x_n)-\zeta_+|\le\delta$ a.s. under $p_\delta(v\mid f)$. So the intensity changes by at most $\delta\sum_pw_{tp}\lambda_p$ whose sum over $t$ is $\delta$. We fix $\delta$ at a small global value (Supplement~S6.2).

\subsection{Evidence lower bound and block updates}
We imitate the latent Poisson components \eqref{eq:superposition} in Section~\ref{sec:prelim}. Conditional on $V$, introduce $H_{tpn}\mid W,\Lambda, V\sim\operatorname{Poisson}(w_{tp}\lambda_pV_{pn}),
~ Y_{tn}=\sum_{p=1}^P H_{tpn}$.
Thus, $H_{tpn}$ is the unobserved contribution of component $p$ to the response at output index $t$ and covariate setting $\bm x_n$. By Poisson superposition, summing these counts recovers the observation model $Y_{tn}\mid V,W,\Lambda\sim\operatorname{Poisson}(\sum_p w_{tp}\lambda_pV_{pn})$. This is similar as \eqref{eq:superposition} with the component weights $\lambda_p$ included.

If $V$ were deterministic and known, conditioning on each observed total would give the classical multinomial allocation in \eqref{eq:classical-allocation}, with probabilities proportional to $w_{tp}\lambda_pV_{pn}$. Here $V$ is random and unobserved, so the latent allocation and amplitude must be inferred jointly.

Write $H=(H_{tpn})$ and
let $\mathcal H(Y)$ contain the nonnegative integer arrays with $\sum_pH_{tpn}=y_{tn}$ for all $t, n$. The smoothed evidence and the variational family for posterior approximation are
\begin{align}
 p_\delta(Y\mid W,\Lambda,\Theta)&=\iint\sum_{H\in\mathcal H(Y)}p(H\mid W,\Lambda, V)p_\delta(V\mid\bm f)p_\Theta(\bm f)\,dV\,d\bm f,\label{eq:evidence}\\
 q(H,V,\bm f)&=q(H)p_\delta(V\mid\bm f)q(\bm f),\quad q(\bm f)=N(\bm\mu,\Sigma),\quad\Sigma\succ0.\label{eq:q}
\end{align}
The approximation $q(H,V,\bm f)$ separates posterior allocations $q(H)$ from $(V,\bm f)$ in a mean field flavor so that its form resembles \eqref{eq:classical-allocation}. The approximation preserves the smoothed conditional amplitude distribution $p_\delta(V\mid\bm f)$ to yield a clean ELBO. No updates for $p_\delta(V\mid\bm f)$. The actual posterior $p(\bm f|Y)$ is not Gaussian but we use a Gaussian $q(\bm f)$ so that taking advantage of the GP structure we can write the posterior predictive distributions at a new $\bm x$ in a clean form (Section \ref{sec:prediction-main}). No mean field approximations across components or settings or latent coordinates for posterior $q(\bm f)$, so that $q(\bm f)$ fits to such posterior dependence. Full posterior covariance $\Sigma$ is retained for $\bm f$. The factor weights $W,\Lambda$ and covariance parameters $\Theta$ are estimated.

In a standard derivation by applying Jensen’s inequality to \eqref{eq:evidence} under $q$, the term $p_\delta(V\mid\bm f)$ cancels by the design in \eqref{eq:q} and the evidence lower bound (ELBO) is
\begin{equation}
\mathcal L:=\E_q\log p(H\mid W,\Lambda, V)-\E_{q(H)}\log q(H)-\KL\{q(\bm f)\Vert p_\Theta(\bm f)\}.
\label{eq:elbo-cancellation}
\end{equation}
We optimize $\mathcal L$ by a four-block alternating algorithm, which we call EMEM. The name follows the variational interpretation of EM \citep{blei2017} and the role of each step, previewed in Algorithm \ref{alg:emem}. E1 and E2 update variational approximations, while M1 and M2 update model parameters. Proposition \ref{prop:updates} shows clear steps.

\begin{algorithm}[t]
\caption{EMEM variational fitting}\label{alg:emem}
\small
\vspace{4pt}
\noindent Initialize feasible $W,\Lambda,\Theta$ and $q(\bm f)=N(\bm\mu,\Sigma)$ with $\Sigma\succ0$. Repeat the following cycle until a stopping criterion or the chosen iteration budget is reached. In each step, hold the other blocks fixed.
\begin{description}[topsep=3pt,partopsep=0pt,parsep=0pt,itemsep=3pt]
\item[E1.] Optimize $q(H)$, i.e., $r_{tpn}$: update the allocations using explicit ratios \eqref{eq:responsibilities}.
\item[M1.] Optimize $W$ and $\Lambda$: update the factorization parameters using \eqref{eq:m1}.
\item[E2.] Optimize $q(\bm f)$, i.e., $\bm\mu$ and $\Sigma\succ0$: fit the Gaussian posterior over the $PN$ latent terms using \eqref{eq:e2}.
\item[M2.] Optimize $\Theta$: estimate the GP prior hyperparameters by minimizing their Gaussian divergence.
\end{description}
\end{algorithm}

\begin{proposition}[Variational EMEM updates]\label{prop:updates}
Fix $\delta>0$ and $\sigma_v^2>0$. Suppose $K_\Theta\succ0$, $\Sigma\succ0$, and every positive observation has a positive fitted intensity. Holding the other blocks fixed within each step, the four updates are as follows.

\smallskip
\noindent\textnormal{\textbf{E1: Allocations.}}
Maximization over $q(H)$ gives $q^*(H_{t\cdot n})=\operatorname{Multinomial}(y_{tn},r_{t\cdot n})$ with,
\begin{equation}\label{eq:responsibilities}
r_{tpn}=\frac{w_{tp}\lambda_p\exp\{\E_q\log V_{pn}\}}{\sum_{j=1}^P w_{tj}\lambda_j\exp\{\E_q\log V_{jn}\}},
\qquad h_{tpn}:=\E_qH_{tpn}=y_{tn}r_{tpn}.
\end{equation}
Zero observations ($y_{tn}=0$) can use arbitrary allocations.

\smallskip
\noindent\textnormal{\textbf{M1: Factorization parameters.}}
Define $A_{tp}=\sum_n h_{tpn}, ~ B_p=\sum_t A_{tp}, ~ a_p=\sum_n\E_qV_{pn}.$
The unique maximizer over $W$ and $\Lambda$, subject to $\sum_t w_{tp}=1$ and $\sum_p\lambda_p=1$, is
\begin{equation}\label{eq:m1}
w_{tp}=\frac{A_{tp}}{B_p},
\quad
\lambda_p=\frac{B_p}{a_p-\eta},
\quad \text{with $\eta$ given by }
\sum_{p=1}^P\frac{B_p}{a_p-\eta}=1,
\quad \eta<\min_p a_p,
\end{equation}
if $B_p>0$ for all components.
The equation for $\eta$ has a unique root in the stated interval.

\smallskip
\noindent\textnormal{\textbf{E2: Gaussian Posterior.}}
Maximization over $q(\bm f)=N(\bm\mu,\Sigma)$ is equivalent to solving
\begin{equation}\label{eq:e2}
\max_{\bm\mu,\,\Sigma\succ0}
\left[
\sum_{p,n}\left\{\left(\sum_t h_{tpn}\right)\E_q\log V_{pn}-\lambda_p\E_qV_{pn}\right\}
-\KL\{N(\bm\mu,\Sigma)\Vert N(0,K_\Theta)\}
\right].
\end{equation}

\smallskip
\noindent\textnormal{\textbf{M2: GP hyperparameters.}}
Maximizing the ELBO over $\Theta$ is equivalent to minimizing $\KL\{N(\bm\mu,\Sigma)\Vert N(0,K_\Theta)\}$
over the admissible GP hyperparameters. In particular, writing $K_\Theta=\sigma_f^2K_0$, the optimal variance for fixed $K_0=K_z\otimes K_x$ is $\sigma_f^2=(\tr(K_0^{-1}\Sigma)+\bm\mu^\top K_0^{-1}\bm\mu) / (PN)$.
\end{proposition}

The derivations and the M1 boundary cases with $B_p=0$ are given in Supplement~S1.3.
The derivations follow the coordinate-ascent algorithm of variational inference \citep[Section~2.4]{blei2017}, so exact maximization of each block gives nondecreasing ELBO values.

The E1 step resembles the form of allocation \eqref{eq:classical-allocation}, with the fixed amplitude replaced by its variational geometric mean $\exp\{\E_q\log V_{pn}\}$.
Although the ratios $r_{tpn}$ have an explicit form, evaluating $\E_q\log V_{pn}$ requires numerical integration. Substituting the optimal multinomial $q(H)$ into \eqref{eq:elbo-cancellation} and suppressing data-only constants gives
\begin{equation}\label{eq:elbo}
\begin{aligned}
\mathcal L
={}&\sum_{t,p,n}
y_{tn}r_{tpn}
\left\{
\log\frac{w_{tp}\lambda_p}{r_{tpn}}
+\E_q\log V_{pn}
\right\}-\sum_{t,p,n}w_{tp}\lambda_p\E_q V_{pn}
-\KL\{q(\bm f)\Vert p_\Theta(\bm f)\}.
\end{aligned}
\end{equation}

M1 then follows these allocations to update the output patterns and component weights. Each column of $W$ is updated to be expected latent counts assigned to that component, normalized across output indices. The update of $\Lambda$ balances component mass $B_p$ to account for its setting-total amplitude $a_p$, reducing weight if $a_p$ is large. The $\Lambda$ update requires only scalar root finding in \eqref{eq:m1}. E1 and M1 therefore avoid numerical optimization over a high dimensional space.

E2 learns the posterior mean and covariance of latent function values. Unlike in M1, the expectations $\E_qV_{pn}$ and $\E_q\log V_{pn}$ now vary with (only) $\mu_{pn}$ and diagonal $\Sigma_{pn,pn}+\sigma_v^2$ from $(\bm\mu,\Sigma)$, while the dependence of all parameters in $(\bm\mu,\Sigma)$ is through the Gaussian divergence. The Poisson part encourages latent amplitudes that explain the allocated observations, and the divergence balances this fit against compatibility with the GP prior. We optimize this deterministic objective using gradients through the numerical integration, with no Monte Carlo estimation or a further lower-bound surrogate as in \cite{basson2023}.
M2 selects the GP hyperparameters that make the latent uncertainty plausible under the prior. The $\sigma_f^2$ update is explicit fixing remaining hyperparameters after they are optimized numerically.

The main computational benefit is that the latent Gaussian has dimension $PN$, independent of the number of output indices $T$, while the GP prior has Kronecker factors of orders $O(P)$ and $O(N)$. This can be particularly useful when each setting produces many outputs but relatively few settings are observed. Because the ELBO \eqref{eq:elbo} separates into factorization and GP divergence terms, the latent GP representation may be extended with more flexible kernels, or with other types of GPs fitted by variational inference. For optimization, we compute gradients of $\E_q\log V_{pn}$ by automatic differentiation through numerical quadrature in experiments (Section \ref{sec:numerical}), showing small numerical error. E2 uses Adam\footnote{Extensive tests of E2 optimizers using second-order information, including trust-region methods and local Gaussian approximations, showed no performance advantage over Adam.}, while M2 uses L-BFGS because its objective is deterministic, smooth, low-dimensional and suitably parameterized. Short inner iterations, capped at 30 E2 steps and 5 M2 updates, were sufficient empirically. We use 15 outer iterations, beyond which predictive error metrics changed little. The amplitude-noise variance $\sigma_v^2$ could also be learned but in preliminary experiments, learning this variance substantially increased optimization time. We therefore keep $\sigma_v^2$ fixed in Proposition~\ref{prop:updates} to reflect our experiment use.

\subsection{A continuous observation model with the same allocation}\label{sec:continuous}

For continuous nonnegative responses, the EMEM updates remain applicable as a Poisson quasi-likelihood, but prediction requires an observation model. We use a mean-preserving jittered Poisson model that spreads each positive latent count over a continuous interval while retaining the atom at zero. It preserves $\E(Y\mid\nu)=\nu$, and Poisson superposition and multinomial allocation remain valid at the latent count level, so its smoothed variational updates are those of Proposition~\ref{prop:updates} with each response replaced by its identified integer count (Supplement~S2.1). Variational inference for continuous responses therefore needs only a rounding step before EMEM. Raw responses give the quasi-likelihood version, which is approximate for this model. In Section~\ref{sec:numerical} it increased the mean and standard deviation of QISE across datasets by 0.55\% and 2.95\%. Other jittered models could also be used.

\subsection{Posterior prediction}\label{sec:prediction-main}
We use $q(\bm f)$ and the conditional structure of GP prior to predict the posterior distribution of the $P$ latent values $\bm f_*$ at a new setting. Let $K_{*f}$ and $K_{**}$ be prior GP covariance matrices related to $\bm f_*$ and $\Pi=K_{*f}K_\Theta^{-1}$ with estimated $\Theta$. Integration over $q(\bm f)$ gives
\begin{equation}\label{eq:prediction}
q_*(\bm f_*)=N(\bm m_*,\Pi_*),\qquad \bm m_*=\Pi\bm\mu,\quad \Pi_*=K_{**}-\Pi K_{f*}+\Pi\Sigma \Pi^\top,
\end{equation}
so the amplitudes before rectification follow $N(\bm m_*,\Pi_*+\sigma_v^2I)$, and with $s_p^2=\Pi_{*,pp}+\sigma_v^2$ the predictive mean is $\sum_pw_{tp}\lambda_p\{m_{*,p}\Phi(m_{*,p}/s_p)+s_p\phi(m_{*,p}/s_p)\}$. Quantiles are numerically approximated by Monte Carlo depending on the specific observation model for $Y$. Partial observations of a new response vector reweight the Monte Carlo draws (Supplement~S5).

\section{Conditional dispersion}\label{sec:dispersion}
\subsection{Controlling conditional dispersion}\label{sec:G}

The Poisson observation model fixes the conditional variance-to-mean ratio at one. This restriction makes the model sensitive to response units. Fitting to rescaled responses does not lead to block optimizers being linearly rescaled in accordance. The restriction arises from the Poisson layer rather than the latent GP representation. In contrast, the algebraic structure of established SVD+GP models (Section \ref{sec:model}) can accommodate such rescaling or change of unit.
We therefore introduce a globally fixed resolution parameter $G$ to improve model flexibility, which controls the inverse conditional dispersion of the observation model.
We replace the observation layer by
\begin{equation}\label{eq:scaled-poisson}
Z_t(\bm x)\mid\bm V(\bm x)\sim\operatorname{Poisson}\{G\nu_t(\bm x)\},
\qquad Y_t(\bm x)=\frac{Z_t(\bm x)}{G}.
\end{equation}
For positive intensity, the conditional variance-to-mean ratio is therefore $1/G$. The original model is recovered at $G=1$. Thus $G$ controls the conditional dispersion while preserving the conditional mean structure. This replacing conditional Poisson model is a special case of the exponential dispersion models in \cite{jorgensen1987}. This redesign applies to the jittered model by jittering $Z$ before dividing by $G$, which also preserves the conditional mean with its variance-to-mean ratio lies between $1/G$ and $69/(64G)$ (Supplement~S2.2).

The parameter $G$ also controls resolution as larger values give finer count units. As a practical choice, we consider $G\in\{1,2,\ldots\}$ to simplify its selection. This restriction does not impose a bound on total predictive variability as $\Var(Y_t(\bm x))=G^{-1}\E(\nu_t(\bm x))+\Var(\nu_t(\bm x))$.
Latent functions can still fit to large response variance as $G$ grows.

Low fractional $G \in (0, 1)$ enlarges coarse jittering intervals, so approximation error can be dominated by misspecifying within-interval density in jittering. This concern also motivates the consideration of high resolution model candidates when a continuous nonnegative response concentrates near zero or when estimating extreme lower quantiles that remain positive. Increasing $G$ narrows jittering intervals and reduces the sensitivity to jittering models, although it does not guarantee better overall predictive accuracy. Section \ref{sec:G-spectrum} illustrates this intuition.

\begin{proposition}[ELBO with $G$]\label{prop:dispersion}
Under a fixed $G$, the posterior allocation in EMEM becomes $q(H_{t\cdot n})=\operatorname{Multinomial}(Gy_{tn},r_{t\cdot n})$ and the resulted ELBO $\mathcal L_G$ is \eqref{eq:elbo} with its first two sums multiplied by $G$.
For fixed remaining blocks, the E1 allocation probabilities and M1 update formulas in Proposition~\ref{prop:updates} remain unchanged, with $h_{tpn}=\E_qH_{tpn}/G=y_{tn}r_{tpn}$.
\end{proposition}

Parameter $G$ scales the factorization terms relative to the Gaussian divergence. As $G$ grows, the ELBO increasingly favours explaining the observations. One set of E1 and M1 update formulas serves a family of models but $G$ still affects fitting posterior in E2 and the divergence objective in M2, and consequently the fitted model. We add subscript $G$ to related notation.

\subsection{A limit connecting joint and separate fitting}\label{sec:largeG}
To better understand the variational inference under large $G$, we let smoothing vanish as $\delta_G=\delta/G$ and optimize $\sigma_v^2\ge0$ (our experiments keep $\sigma_v^2>0$ fixed).

\begin{theorem}[Leading factorization limit]\label{prop:largeG}
Let $Y$ have finite nonnegative entries and positive total. Let $\mathcal K$ be a fixed nonempty family of positive definite GP covariance matrices and $\mathcal L_G^*$ the supremum of ELBO $\mathcal L_G$ in Proposition \ref{prop:dispersion} over feasible variables. Let $L_Y(\cdot, \cdot)$ denote the two-factor KL-NMF objective for matrix $Y$ and $L_Y^*=\sup_{W,V}L_Y(W,V)$ subject to column normalization for $W$. Any finite feasible EMEM solution sequence with $0\le\mathcal L_G^*-\mathcal L_G=o(G)$ has, for $V_G:=\E_{q, G}V$,
\begin{equation}\label{eq:largeG-leading}
\mathcal L_G/G\longrightarrow L_Y^*,\qquad
L_Y(W_G,\Lambda_GV_G)\longrightarrow L_Y^*.
\end{equation}
As $G\to\infty$, every sequence of factor pairs $(W_G,\Lambda_GV_G)$ has a subsequence converging to a global KL-NMF solution.
\end{theorem}
The factorization term in ELBO $\mathcal L_G$ (Proposition \ref{prop:dispersion}) does not determine the GP fit with a divergence objective term of order one. For $q_G (\bm f)=N(\bm\mu_G,\Sigma_G)$, write the M2 objective, up to a term independent of $K$, as $\mathcal L^{\mathrm {GP}}_G(K) :=-\tfrac12\log|K|-\tfrac12\tr\{K^{-1}(\Sigma_G+\bm\mu_G\bm\mu_G^\top)\}.$

\begin{theorem}[GP fitting to a latent completion]\label{prop:completion}
Under Theorem~\ref{prop:largeG}, suppose $\mathcal K$ is closed under positive rescaling and the covariance-block gap satisfies $0\le\sup_{K\in\mathcal K}\mathcal L^{\mathrm {GP}}_G(K)-\mathcal L^{\mathrm {GP}}_G(K_G)=o(1)$. Consider a solution subsequence such that $(W_G,\Lambda_G,V_G,K_G)\to(W_\infty,\Lambda_\infty,V_\infty,K_\infty)$, every limiting component weight is positive, and $K_\infty\in\mathcal K$ is positive definite. Then $\Sigma_G\to0$, $\sigma_{v,G}^2\to0$, and sequence $\{\bm\mu_G\}$ is bounded. Every point $\bm f^\dagger$ as a subsequence limit of $\{\bm\mu_G\}$ satisfies
\begin{equation}\label{eq:largeG-gp}
(\bm f^\dagger)_+=\operatorname{vec}(V_\infty),\qquad
K_\infty\in\argmax_{K\in\mathcal K}\log\phi_K(\bm f^\dagger),
\end{equation}
where $\phi_K$ is the zero-mean Gaussian density with covariance $K$.
\end{theorem}
Thus a limiting joint fit combines a global factorization with noiseless GP likelihood fitting to a compatible latent completion $\bm f^\dagger$. Positive amplitudes are retained while zero amplitudes permit nonpositive completed values. If every amplitude is positive, the result reduces to GP fitting directly to the right factor. The completion is selected through the Gaussian variational fit and need not be unique. For clarification, the result does not identify an arbitrary independent KL-NMF fit followed by a user-prescribed completion rule (e.g., for completing Figure~\ref{fig:nmf}). As we set as a benchmark in Section~\ref{sec:numerical}, KL-NMF followed by censored GP regression is a distinct method/model. Proofs of above theorem results are in Supplement~S3.

\subsection{Complementary shape estimates}\label{sec:G-spectrum}

Varying $G$ gives a spectrum of EMEM fits, with the separate fitting interpretation in Section~\ref{sec:largeG} at its large-$G$ end. As motivated in Section~\ref{sec:model}, separate fitting can be restrictive because a factorization that best reconstructs the observations need not provide latent amplitudes that are best predicted across covariate settings. The opposite end of the spectrum also has limitations. Let $Q_G$ be a fitted predictive quantile function with finite mean $m_G$ bounded away from zero, and let the true quantile function satisfy $Q_*\in L^2(0,1)$. For both Poisson and continuous responses, supplementary Lemma~S2 gives $\|Q_G-Q_*\|_2
\geq
\left\{\sqrt{m_G/G}-\|Q_*\|_2\right\}_+.$
Integrated squared quantile error diverges as $G\downarrow0$ even when the fitted mean is accurate. For continuous responses, jittering creates further misspecification error which can dominate predictive accuracy at low resolution, including fractional $G<1$. These considerations motivate us to seek an optimal model between the two ends.

\begin{figure}[!htbp]
\centering
\includegraphics[width=0.8\linewidth]{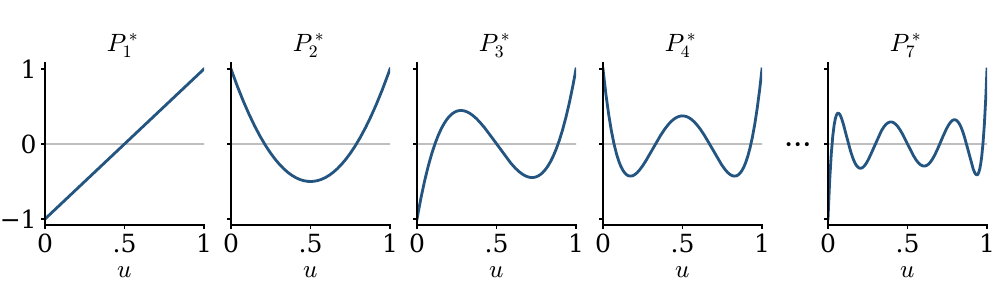}
\caption{Shifted Legendre polynomials of degrees $1, \dots, 4,$ and $7$. The $r$th L-moment uses $P_{r-1}^*$, so the first eight L-moments use the constant $P_0^*=1$ and degrees up to 7.}
\label{fig:legendre}
\end{figure}

Instead of requiring one optimal $G$ to describe every aspect of a predictive distribution, we investigate whether different fits capture complementary shape features. Among possible distributional summaries, L-moments provide well-studied characteristics through orthogonal functions of quantile level \citep{hosking1990}. Let $P_j^*(u)=P_j(2u-1)$ shifted from the usual Legendre polynomial $P_j$ of degree $j$. They satisfy orthogonality $\int_0^1 P_j^*(u)P_k^*(u)\,du
=\frac{\mathbf 1\{j=k\}}{2j+1}$ (Figure~\ref{fig:legendre}).
For a quantile function $Q$, define L-moments and their ratios as
\begin{equation}\label{eq:lmom}
L_r(Q)=\int_0^1 Q(u)P_{r-1}^*(u)\,du,\quad r\geq1,
\quad \text{and} \quad
\tau_r(Q)=\frac{L_r(Q)}{L_2(Q)},\quad r\geq3,
\end{equation}
where the ratios require $L_2(Q)>0$. Here $L_1$ is the mean, $L_2$ measures dispersion, and $\tau_3$ and $\tau_4$ are L-skewness and L-kurtosis. The ratios remove location and scale, allowing comparison of distributional shape. Every L-moment exists under a finite absolute first moment, accommodating discrete, continuous and mixed distributions.

Orthogonality makes the L-moments describe distinct shape features of quantile function. Higher degrees introduce finer oscillations near the endpoints (Figure~\ref{fig:legendre}). Higher-order L-moments encode finer shape information relevant to extreme tail events. The exact-moment construction in Section~\ref{sec:diagnostics} shows that even the exact first eight L-moments give small integrated error while leaving substantial error at extreme quantiles.

\begin{figure}[!htbp]
\centering
\includegraphics[width=0.75\linewidth]{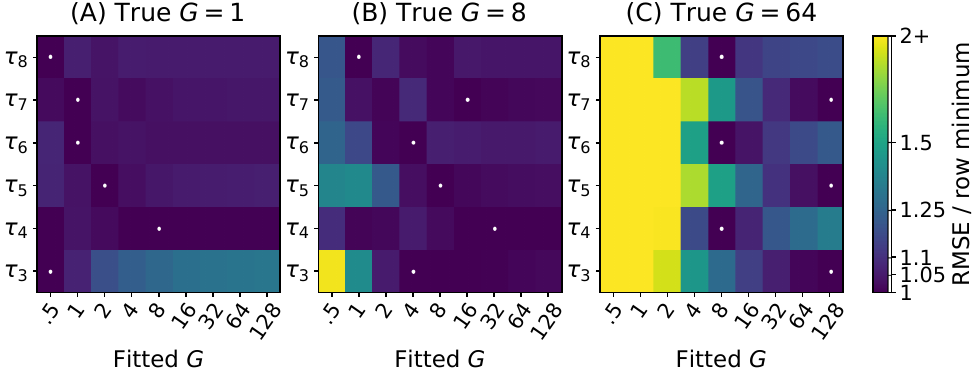}
\caption{Relative RMSE of fitted L-moment ratios for three generating values of $G$. Colors compare candidate fits within a row, not absolute errors across orders or subplots.}
\label{fig:complementarity}
\end{figure}

In a motivating experiment, responses are generated from ten rectified GP factors and our jittered observation model with true $G\in\{1,8,64\}$ (Supplement~S6.1). We fit candidates at $\mathcal G=\{0.5,1,\dots,64,128\}$ under equal computational budget. Let $\RMSE_r(G)$ be the root mean squared error of $\widehat\tau_{r,G}$, pooling squared errors over response entries for four datasets, and $\Rel_r(G)=\RMSE_r(G)/\min_{g\in\mathcal G}\RMSE_r(g)$ (Figure~\ref{fig:complementarity}). The clearest complementarity occurs at true $G=64$. Fitted $G=128$ minimizes the errors of $\tau_3,\tau_5,\tau_7$ and $G=8$ those of $\tau_4,\tau_6,\tau_8$. The best single model, $G=32$, which minimizes the largest $\Rel_r(G)$, is still at least 21\% above the order-specific minimum for any ratio (respectively 34\% and 18\% for the two synthesized latent functions). The true $G=1$ and $8$ settings give smaller compromise. Some optima are interior and others lie at the grid boundary.

This observation motivates combining information from several fitted distributions so that each contributes the shape features it estimates most accurately. Section~\ref{sec:quantile} develops such a construction by selecting L-moments, or their ratios, separately and combining them through orthogonal quantile expansion.

\section{Quantile construction}\label{sec:quantile}
\subsection{Combining L-moments from different fits}
An orthogonal expansion lets different predictive distributions, from EMEM or any other model, supply different shape features. The expansion is based on \citet{sillitto1969} and stated in L-moment form by \citet[Theorem 3]{hosking1990}. For a quantile function $Q\in L^2(0,1)$, let $Q_R(u)=\sum_{r=1}^R(2r-1)L_r(Q)P_{r-1}^*(u)$. Then $\|Q-Q_R\|_2^2=\sum_{r>R}(2r-1)L_r(Q)^2\longrightarrow0$, as $R\to\infty$.\footnote{The same mean-square expansion applies to distributions with probability masses because the shifted Legendre polynomials form a complete orthogonal basis of \(L^2(0,1)\).}
Given supplied L-moments $\widetilde L_1,\ldots,\widetilde L_R$, we construct
\begin{equation}\label{eq:construction}
\widetilde Q_R(u)=\sum_{r=1}^R(2r-1)\widetilde L_rP_{r-1}^*(u),\qquad \widehat Q_R=\mathcal R\{\pos{\widetilde Q_R}\},
\end{equation}
where operation $\mathcal R$ is the rearrangement in \cite{chernozhukov2010}. Clipping imposes the known nonnegativity and rearrangement restores monotonicity, so $\widehat Q_R$ is a valid nonnegative quantile function (repaired from $\widetilde Q_R$ as a finite polynomial that is not necessarily monotone or nonnegative). Numerical checks show that nonmonotonicity occurred mainly in the lower tail.

\begin{proposition}\label{prop:repair}
If the true quantile function $Q_*\in L^2(0,1)$ is nonnegative, then
\begin{equation}\label{eq:repair}
\|\widehat Q_R-Q_*\|_2^2\le\sum_{r=1}^R(2r-1)\{\widetilde L_r-L_r(Q_*)\}^2+\sum_{r>R}(2r-1)L_r(Q_*)^2.
\end{equation}
\end{proposition}
Orthogonality separates L-moment error from truncation error, and neither clipping nor rearrangement can increase the integrated squared error (Supplement~S4.1). The bound justifies choosing each L-moment by its own error. It also describes the tradeoff in selecting $R$ as high $R$ reduces truncation error but requires estimating L-moment of higher orders. So retaining one candidate's expansion residual beyond order $R$ does not necessarily improve accuracy (confirmed by our preliminary numerical checks).
The choice of $R$ also depends on the prediction target. Small integrated error does not guarantee accurate extreme quantiles, which may require information beyond the retained orders. This distinction motivates the diagnostics in Section~\ref{sec:diagnostics}.

The first L-moment equals the mean, so a trusted predictor $\widehat m(\bm x,t)$ can supply $\widetilde L_1$. The other L-moments then describe dispersion and shape around that prediction. Before repairing, changing $\widetilde L_1$ simply shifts the quantile curve and leaves higher L-moments unchanged. It contributes exactly its mean square error $(\widehat m-L_{1,*})^2$ to the bound \eqref{eq:repair}. Thus the construction can use any trusted external point predictor or simulator (itself need not be a GP) while obtaining fitted distributional shape from our nonnegativity-respecting models.

\subsection{Validation scores for higher L-moments}\label{sec:moment-score}
Selection is hard in practice because the true L-moments are unknown and each unique pair $(\bm x,t)$ may carry only one observation. We therefore need validation scores computable from single held-out responses. At a fixed $(\bm x,t)$, let $F_*,Q_*$ be the true distribution and quantile functions, with L-moments $L_{r,*}=L_r(Q_*)$, and let $F_0,Q_0$ describe a common reference distribution. Candidates and reference are fitted without the validation vector containing $Y_\mathrm{v}\in \R_+$. All expectations below condition on that fitting and the $(\bm x,t)$ entry. Reference selection is discussed later.

Write $p_r(u)=P_{r-1}^*(u)$. Specializing the transformed response of \citet[Section~3.2, equation~(13)]{chetverikov2025} to the weight $p_r$ gives the pseudo-observation
\begin{equation}\label{eq:pseudo}
Z_r(Y_\mathrm{v};Q_0)
=L_r(Q_0)+\int_{\R_+} p_r(F_0(y))
\{F_0(y)-\ind(Y_\mathrm{v}\le y)\}\,dy,
\qquad r\ge2.
\end{equation}

The reference provides a starting estimate of the true L-moment, $L_r(Q_0)$, and the held-out response supports its influence-function correction. Thus a single response contributes information about the reference error without requiring repeated responses at the same $(\bm x,t)$ pair.
For L-moments, the correction gives the following bounds.

\begin{proposition}[Bounded sensitivity]\label{prop:sensitivity}
Assume $\E|Y_\mathrm{v}|<\infty$ and $Q_0\in L^1(0,1)$. For $r\ge2$, let $e_r=\E Z_r-L_{r,*}$ denote the mean bias and $d_F=\int_\R(F_0(y)-F_*(y))^2\,dy$. Then
\begin{equation}\label{eq:bias-bound}
e_2=d_F,\qquad |e_r|\le\frac{r(r-1)}2d_F, \qquad
d_F\le\|F_0-F_*\|_\infty\|Q_0-Q_*\|_1.
\end{equation}
For fixed $Q_0$, $|Z_r(y;Q_0)-Z_r(y';Q_0)|\le|y-y'|$.
If $\E Y_\mathrm{v}^2<\infty$, then $\Var(Z_r)\le\Var(Y_\mathrm{v})$.
\end{proposition}

These results also hold for distributions with probability masses. For a finite candidate L-moment $a$, define score
\begin{equation}\label{eq:lm-score}
S_r(a,Y_\mathrm{v})=a^2-2aZ_r.
\end{equation}
This differs from $(a-Z_r)^2$ by the term $Z_r^2$, which is common to all candidates using the same reference, so they give identical empirical selections. Under the assumptions of Proposition~\ref{prop:sensitivity}, $\E S_r(a,Y_\mathrm{v})
=(a-L_{r,*})^2-L_{r,*}^2-2ae_r.$
When the reference is correct, $e_r=0$, so expected scores rank candidates by squared L-moment error. This comparison requires only a finite first moment, thus suitable for the black-box problem.

The bounded sensitivity is seen in multiple aspects. If using the reference L-moment directly, removing the correction part gives $|L_r(Q_0)-L_{r,*}|\le\|Q_0-Q_*\|_1$, whereas the mean-error bound \eqref{eq:bias-bound} shows an additional factor $\|F_0-F_*\|_\infty$ that is small when $F_0$ is uniformly close to the truth. Although the error bound grows with $r$, the $Y_\mathrm{v}$-corrected estimate is 1-Lipschitz in $Y_\mathrm{v}$ at every order. Along with correction, $Y_\mathrm{v}$ introduces response randomness to the pseudo-observation, but Proposition \ref{prop:sensitivity} also shows that its variance does not exceed the response variance at every order.

For two finite candidate moments $a,b$, $\E[S_r(a,Y_\mathrm{v})-S_r(b,Y_\mathrm{v})]
=(a-L_{r,*})^2-(b-L_{r,*})^2-2(a-b)e_r.$
A strict ordering by squared error is therefore preserved in expectation whenever
\[
\left|(a-L_{r,*})^2-(b-L_{r,*})^2\right|
>r(r-1)|a-b|d_F.
\]
Proposition \ref{prop:sensitivity} then implies that correction reduces the impact of reference error in expected comparisons.
High-order selection can be difficult because candidate moments may be close and reference bias can dominate their differences. But we may not need correct selection when candidates give similar moment errors, e.g., when $|(b-L_{r,*})^2-(a-L_{r,*})^2|\le r(r-1)|a-b|d_F$.

At order two, the pseudo-observation is the continuous ranked probability score (CRPS) of the reference distribution \citep{gneiting2007}. Formally,
\begin{equation}\label{eq:Z2-crps}
Z_2(Y_{\mathrm v};Q_0)=\operatorname{CRPS}(F_0,Y_{\mathrm v}):=\int_{\mathbb R}(F_0(z)-\mathbf{1}(Y_{\mathrm v}\le z))^2\,dz,\qquad \E Z_2=L_{2,*}+d_F.
\end{equation}
Thus minimizing expected CRPS minimizes $d_F$ and the upper bound in \eqref{eq:bias-bound} for every order, which remains true if using the same validation responses for all reference candidates. This motivates selecting the reference by empirical CRPS within the training data.
This criterion also requires only finite first moments

We use nested cross-validation (CV) over whole response vectors. Within each outer training set, inner validation selects the reference from EMEM fits over a set $\mathcal G$ of $G$ values by average CRPS.
The selected reference and all moment candidates are refitted on the outer training set. The reference supplies a common $Z_r$ for all candidates at each outer validation response entry. Write $S_r^{(-k)}$ for \eqref{eq:lm-score} using the reference fitted without outer fold $I_k$. We select one candidate by
\begin{equation}\label{eq:cv}
\widehat G_r
=\argmin_{G\in\mathcal G}
\frac1K\sum_{k=1}^K
\frac1{|I_k|T}\sum_{n\in I_k}\sum_{t=1}^T
S_r^{(-k)}
\big( \widehat L_{r,G}^{(-k)}(\bm x_n,t),y_{tn}\big)
\end{equation}
for each order. The selected candidates are then refitted on all training vectors, and their selected L-moments enter \eqref{eq:construction}.

\subsection{Selecting L-moment ratios}\label{sec:ratio-score}

To exploit the complementarity seen in Figure~\ref{fig:complementarity}, we can select $L_2$ as above and then select L-moment ratios for higher orders. For a candidate with $\widehat L_{2,G}>0$, suppose $L_{2,*}>0$ and define
\begin{equation}\label{eq:ratio-score}
t_{r,G}=\frac{\widehat L_{r,G}}{\widehat L_{2,G}},
\qquad
S_r^{\rm R}(t_{r,G},Y_\mathrm{v})
=t_{r,G}^2Z_2-2t_{r,G}Z_r,
\qquad 3\le r\le R.
\end{equation}
At a fixed entry and conditional on the fitted candidate, this applies the transformation in \eqref{eq:pseudo} to $t_{r,G}^2L_2-2t_{r,G}L_r$, with weight $t_{r,G}^2p_2-2t_{r,G}p_r$. Taking expectations gives
\begin{equation}\label{eq:weighted-risk}
\E S_r^{\rm R}(t_{r,G},Y_\mathrm{v})
=L_{2,*}\big((t_{r,G}-\tau_{r,*})^2-\tau_{r,*}^2\big)
+t_{r,G}^2e_2-2t_{r,G}e_r.
\end{equation}
Similar to L-moment selection, under a correct reference, the expected score $\E S_r^{\rm R}$ ranks squared ratio errors. Averaging over validation entries targets $L_{2,*}$-weighted squared ratio error. The score avoids division by the reference's $L_2$.
After refitting the selected candidates, provide $\widetilde L_1=\widehat m$, $\widetilde L_2=\widehat L_{2,\widehat G_2}$, and $\widetilde L_r
=\widetilde L_2
{\widehat L_{r,\widehat G_r}}/
{\widehat L_{2,\widehat G_r}}$, $3\le r\le R$, for the construction \eqref{eq:construction}.
We refer to the construction method with selection of L-moments as LM-CV and with this weighted ratio selection as R-CV.

\section{Numerical experiments}\label{sec:numerical}
\subsection{Design and methods}\label{sec:synthetic_exp_design}
We test whether quantile construction helps in a difficult but normal case where responses follow a distribution outside the fitted family of every candidate model. For $\bm x\in[0.1,2]^3$ and $t\ge0$, let
\begin{align}\label{eq:base}
b(\bm x,t)={}&\left[e^{-0.01tx_1}+e^{-0.005tx_2}+(1+0.1tx_3)^{-1}\right]\nonumber\\[-2pt]
&\cdot[1+\sin(0.1tx_1)][1+\sin(0.2tx_2+3\pi/2)][1+\cos(0.15tx_3)]
\end{align}
be a base function and let $m=b+0.01$ for strict positivity and numerical stability of true extreme lower quantiles. Generate responses independently as $Y_t(\bm x)=m+U$ with $U\sim\operatorname{Gamma}\left(\text{shape}=\frac{(m+6)^2}{3m},\ \text{scale}=\frac{3m}{m+6}\right)$,
implying $\E Y=2m+6$ and $\Var(Y)=3m$. Seemingly unpredictable peaks of the base function (Figure~\ref{fig:base}) challenge mean prediction and give scale, skewness, and support boundary $m$ that all vary with the response index and covariates. The boundary $m$ is not exactly identified by the knowledge of nonnegativity but is close to zero.

Ten independently regenerated response datasets share 50 Latin-hypercube training $\bm x$'s, ten uniform test $\bm x$'s and $T=1000$ output indices $t$ on $[0,100]$. Methods share data and folds within each dataset. EMEM uses $P=10$, $d_z=2$, amplitude-noise variance $\sigma_v^2=\delta^2/4$, one initialization, $G\in\{1,2,3,6,9,12,24,48,96,192,384\}$, and our jittered model.
Constructions use $R=8$ and the mean uses a wavelet implicit neural representation (WIRE) fit \citep{saragadam2023}. Each native EMEM reference quantile is shifted to match the WIRE mean, then also clipped and rearranged. Five outer folds and two inner validation rounds per outer fold select references by CRPS and moments by LM-CV or ratios by R-CV. All numerical evaluations meet the 0.1\% numerical error target.

Three rank-ten standalone benchmarks are GP regression on SVD's right component (SVD+GP, \citealp{higdon2008}), KL-NMF followed by the censored GP regression in \cite{basson2023}, and GPP-NMF \citep{schmidt2008}. We extend GPP-NMF to predict its right-factor GP conditional on fitted left factors and MAP parameters, retaining its Gaussian noise model. Gaussian-response predictions are clipped at zero.
For $N_*=10$ test settings, define the quantile integrated squared error (QISE) and distribution-function integrated squared error (CDFISE) by
\begin{align}\label{eq:metrics}
\operatorname{QISE}=\frac1{N_*T}\sum_{n,t}\int_0^1(\widehat Q_{tn}-Q_{*,tn})^2\,du, \quad
\operatorname{CDFISE}=\frac1{N_*T}\sum_{n,t}\int_{\R}(\widehat F_{tn}-F_{*,tn})^2\,dy.
\end{align}
Quantile, L-moment and L-moment ratio RMSEs are also computed within each dataset over its $N_*T=10,000$ test locations. All metrics are averaged across the ten datasets.
\subsection{Experiment results}

Relative to respective best single-$G$ candidate, R-CV reduces QISE by 31.1\% and CDFISE by 19.4\% (Table~\ref{tab:main}), indicating construction effectiveness. The QISE improvement is consistent with the theory motivation. As tested in Section \ref{sec:diagnostics}, using exact moments gives low ISEs and quantile RMSEs at intermediate levels but does not ensure equally accurate extreme quantiles, which remain a more demanding objective. QISE integrates errors over probability levels, so errors related to a narrow tail can contribute little, particularly for lower tails near zero. Indeed, LM-CV gives the smallest quantile errors at 0.9999 and nearly at 0.0001.
Benchmark strengths also depend on the error metrics. Clipped SVD+GP has the smallest benchmark integrated and upper-tail errors, while KL-NMF + censored GP is most accurate at the lower quantiles.

\begin{table}[!htbp]
\centering
\setlength{\tabcolsep}{3pt}
\begin{tabular}{@{}lrrrrrrrr@{}}
\toprule
& & & \multicolumn{6}{c}{Quantile RMSE}\\
Method & QISE & CDFISE & $0.0001$ & $0.001$ & $0.01$ & $0.99$ & $0.999$ & $0.9999$\\
\midrule
EMEM, $G=1$ & 35.5 & 1.81 & 6.08 & 6.19 & 5.97 & 11.3 & 14.6 & 17.6\\
EMEM, $G=12$ & 31.5 & 1.97 & 4.92 & 4.88 & 4.70 & 9.88 & 12.2 & 14.4\\
EMEM, $G=96$ & 31.8 & 2.02 & 4.90 & 4.85 & 4.67 & 9.90 & 12.2 & 14.4\\
EMEM + LM-CV & 21.8 & 1.46 & 4.11 & 4.24 & 4.41 & 8.44 & 9.57 & 10.9\\
EMEM + R-CV & 21.7 & 1.46 & 4.25 & 4.26 & 4.34 & 8.22 & 9.54 & 11.2\\
\addlinespace[3pt]
KL-NMF + censored GP & 31.9 & 1.82 & 4.09 & 4.18 & 4.20 & 11.2 & 14.0 & 16.4\\
SVD + GP, clipped & 23.9 & 1.54 & 4.81 & 4.90 & 4.77 & 8.66 & 10.9 & 13.0\\
GPP-NMF, clipped & 29.9 & 1.78 & 5.07 & 5.31 & 5.47 & 9.20 & 11.3 & 13.3\\
\bottomrule
\end{tabular}
\caption{Average test errors across ten response datasets, to three significant figures. Smaller values are better. Constructions use eight L-moments, the WIRE point predictor and CRPS-selected references. Only three of the eleven candidate \(G\) values are displayed. }\label{tab:main}
\end{table}

At probabilities 0.0001, 0.001 and 0.01, R-CV reduces RMSE by 13.3\% (16.0\% for LM-CV), 12.2\% and 7.2\% relative to respective best single-$G$ candidate. Both constructions outperform clipped SVD+GP and GPP-NMF\footnote{Clipping directly reduces the RMSE at 0.0001 of SVD+GP and GPP-NMF by 43.4\% and 47.7\% respectively.} at these levels, remaining only slightly less accurate than KL-NMF+censored GP. Predicting small lower quantiles under our construction criteria can be difficult. The response lower tail is bounded and close to zero at low probability levels but its upper tail can be unbounded. Errors of comparable size can have greater influence on L-moments through upper quantiles, since shifted Legendre polynomials are symmetric. Therefore selection guided by L-moment does not ensure small lower tail errors. On the other hand, an overly dispersed predictive distribution can easily produce small lower quantile errors (if the quantile is close to the bound) while predicted upper quantiles exceeding the truth.
Although extreme lower quantiles remain difficult, their closeness to zero limits absolute errors. The unbounded upper tail permits much larger errors, making the upper-tail improvements particularly important.
At probabilities 0.99, 0.999 and 0.9999, R-CV reduces RMSE by 16.7\%, 21.9\% and 22.1\% (23.6\% for LM-CV) relative to respective best single-$G$ candidate. The reductions relative to SVD+GP, the strongest benchmark at these levels, are 5.1\%, 12.3\% and 13.8\% (15.6\%).
Both constructions outperform every standalone benchmark at these upper levels.
The upper-tail gains better reflect the advantage of quantile construction.

With CRPS-selected references, the shared $L_2$ (order-two L-moment) score identifies the correct candidate in all ten datasets for both constructions. This $L_2$ correctness is particularly notable because $L_1$ is supplied by WIRE and candidate L-moment RMSEs differ most at order two.\footnote{Across $G$, the mean $L_2$ RMSE ranges from 1.203 to 1.674, whereas the range at each order $r=3,\ldots,8$ is at most 0.064.}
Selecting L-moment ratios does improve all six final ratio RMSEs over LM-CV, with high reductions of approximately 86\%, 84\% and 79\% for $\tau_4$, $\tau_6$ and $\tau_8$.
\subsection{Sources of the remaining error}\label{sec:diagnostics}
\noindent \textbf{Truncation.} Low-order L-moment accuracy does not ensure accurate extreme quantiles. The construction from the exact first eight L-moments has QISE 0.0317, but quantile RMSEs of 1.05 and 5.34 at probabilities 0.0001 and 0.9999, compared with 0.0683 and 0.167 at less extreme levels 0.1 and 0.9. Relative to the best candidate, R-CV closes 35.2\% (LM-CV 37.7\%) of the RMSE gap to this exact-moment construction at 0.9999. Higher orders may reduce truncation error but require more estimation and selection accuracy.

\noindent \textbf{Mean.} Mean prediction error can dominate QISE. For a repaired $\widehat Q$ with mean $\widehat m$ and true mean $m_*$, $\|\widehat Q-Q_*\|_2^2=(\widehat m-m_*)^2+\|(\widehat Q-\widehat m)-(Q_*-m_*)\|_2^2$ by orthogonality. Replacing the mean by WIRE in the exact-8-moment construction raises average QISE to 17.39. Relative to the best candidate, LM-CV and R-CV close 68.9\% and 69.3\% of the QISE gap respectively.

\noindent \textbf{Reference and selection.} Replacing the CRPS-selected reference by the true quantile reduces LM-CV QISE from 21.79 to 21.69. This supports the reference sensitivity discussion following Proposition \ref{prop:sensitivity}.

\subsection{A hydrological risk demonstration}\label{sec:demo}
Agricultural production faces hydrological risk at both ends of the streamflow distribution: low flows constrain irrigation water, while high flows accompany flooding that damages cropland and infrastructure. We use observed daily streamflow and ERA5 meteorological variables from the 2023 update of HYSETS \citep{arsenault2020} for 42 cropland-dominated Iowa catchments with complete records over 1981--2023. Each year is divided into four 90-day seasons, and concatenating daily runoff across catchments gives $T=3{,}780$ responses for each of 172 seasons. The four covariates are regional averages of seasonal precipitation, the fraction of days with precipitation below 1 mm/day, downward shortwave radiation, and precipitation over the preceding 90 days, so predictions are conditional on regional seasonal weather.

We use 1981--2006 for initial fitting, 2007--2010 for reference selection, 2011--2014 for L-moment selection and 2015--2023 for testing, without rotation. Method settings receive no tuning. Selected candidates are refitted on all $N=136$ seasons through 2014. To equip an existing point simulator with predictive intervals, we take raw daily discharge from the separate GEOGLOWS v2 simulation,\footnote{GEOGLOWS \citep{hales2025advancing} uses time-varying ERA5 runoff inputs and models water transport through the river network, incorporating temporal and spatial information beyond the four regional seasonal weather summaries supplied to EMEM and SVD+GP. \url{https://training.geoglows.org/rfs/datasets/retrospective/}.} and convert it to runoff depth to supply $\widetilde L_1$.  Geographic matching retains 35 catchments. We evaluate central entrywise 90\% and 99\% intervals by coverage (in \%), lower and upper miss rates (in \%), width (in mm/day), and the interval score (in mm/day) of \citet{gneiting2007}, all averaged over test catchment-days (Table~\ref{tab:hydrology}).

\begin{table}[!htbp]
\centering
\setlength{\tabcolsep}{2.5pt}
\begin{tabular}{@{}lrrrrrrrrrr@{}}
\toprule
& \multicolumn{5}{c}{90\% interval} & \multicolumn{5}{c}{99\% interval}\\
\cmidrule(lr){2-6}\cmidrule(l){7-11}
Method & Cov. & Lower & Upper & Width & Score & Cov. & Lower & Upper & Width & Score\\
\midrule
SVD+GP, clipped          & 83.2 &  5.01 & 11.80 & 1.63 & 5.30 & 92.1 &  0.88 & 7.01 & 2.31 & 27.10\\
LM-CV + EMEM $L_1$       & 80.6 & 15.57 &  3.83 & 2.51 & 5.35 & 84.3 & 14.27 & 1.45 & 4.23 & 23.32\\
\quad + GEOGLOWS $L_1$   & 93.1 &  2.06 &  4.86 & 2.05 & 4.43 & 96.6 &  2.00 & 1.45 & 3.76 & 18.07\\
R-CV + EMEM $L_1$        & 65.2 & 29.41 &  5.38 & 2.09 & 5.50 & 81.8 & 15.48 & 2.69 & 3.18 & 24.90\\
\quad + GEOGLOWS $L_1$   & 90.1 &  2.52 &  7.34 & 1.69 & 4.53 & 94.9 &  2.07 & 3.07 & 2.70 & 21.86\\
\bottomrule
\end{tabular}
\caption{Test results on 35 matched catchments (113,400 catchment-days). Nominal miss rates are 5 at 90\% and 0.5 at 99\% per side. A point mass at zero can make a calibrated lower-miss rate smaller than nominal because misses use strict inequalities.}\label{tab:hydrology}
\end{table}

The constructions with their own first moments miss low flows with rates of 14.27--29.41\%. Supplying the point simulation reduces lower-miss rates by 86.0--91.4\% and closes on average 81.5\% of the absolute gap between observed and nominal coverage, removing most of the lower-tail deficiency. Interval scores fall by 17.2\% and 22.5\% for LM-CV and by 17.5\% and 12.2\% for R-CV at 90\% and 99\%. These reductions and coverage increases persist when any single test year is excluded.
Both simulator-assisted constructions now have lower interval scores than SVD+GP at both levels by 14.5--33.3\%, while SVD+GP excessively misses high flows.

\subsection{Discussion}\label{sec:discussion}
Knowledge of a response lower bound should shape its predictive distribution. In this work, the latent GP factor model respects nonnegativity throughout and, across a spectrum of conditional dispersion, supplies candidates that capture different shape features of the truth, with no single dispersion preferred for all orders. Selection is guided by L-moment errors and the integrated quantile error into which they combine. When extreme lower quantiles and their predictions lie close to zero, unbounded upper-tail errors can have greater influence on these criteria. Uncertainty near the lower bound motivates model design, while upper-tail uncertainty guides selection: knowledge of the lower bound shapes our quantification of upper-tail uncertainty.

\section*{Acknowledgements}
Part of this research was performed while the author was visiting the Institute for Mathematical and Statistical Innovation (IMSI), supported by the National Science Foundation (Grant No.\ DMS-1929348).
Generative artificial intelligence tools assisted with coding, writing, and mathematical derivations. The author is responsible for the final content.

\begingroup
\setstretch{1.6551724}\normalsize
\setlength{\bibsep}{2pt}\interlinepenalty=10000
\bibliographystyle{ememnat}
\bibliography{references}
\endgroup
\end{document}